\RequirePackage{ifpdf}
\ifpdf 
\documentclass[pdftex]{sigma}
\else
\documentclass{sigma}
\fi

\DeclareMathOperator{\Lin}{Lin}

\begin{document}
\allowdisplaybreaks

\newcommand{\by}[1]{\textit{{#1}}}
\newcommand{\jour}[1]{\textit{{#1}}}
\newcommand{\vol}[1]{\textbf{V.{#1}}}
\newcommand{\book}[1]{\textrm{{#1}}}

\newcommand{\Id}{{\mathrm d}}
\newcommand{\BBC}{{\mathbb{C}}}
\newcommand{\BBR}{{\mathbb{R}}}
\newcommand{\BBN}{{\mathbb{N}}}
\newcommand{\sym}{\mathop{\rm sym}\nolimits}
\newcommand{\const}{\mathop{\rm const}\nolimits}

\newcommand{\gothg}{\mathfrak{g}}
\newcommand{\gm}{\mathfrak{m}}
\newcommand{\bgm}{{\bar{\mathfrak{m}}}}
\newcommand{\gt}{\mathfrak{t}}
\newcommand{\bi}{{\boldsymbol{i}}}
\newcommand{\bu}{{\boldsymbol{u}}}
\newcommand{\bU}{{\boldsymbol{U}}}
\newcommand{\ba}{{\boldsymbol{a}}}
\newcommand{\balpha}{{\boldsymbol{\alpha}}}
\newcommand{\bx}{{\boldsymbol{x}}}
\newcommand{\bE}{\mathbf{E}}
\newcommand{\bT}{\mathbf{T}}
\newcommand{\bw}{{\boldsymbol{W}}}
\newcommand{\gA}{\mathfrak{A}}
\newcommand{\gB}{\mathfrak{B}}
\newcommand{\gC}{\mathfrak{C}}
\newcommand{\sA}{\mathsf{A}}
\newcommand{\cE}{\mathcal{E}}
\newcommand{\cEEL}{{\mathcal{E}}_{\text{\textup{EL}}}}
\newcommand{\cH}{\mathcal{H}}
\newcommand{\cL}{\mathcal{L}}
\newcommand{\cR}{\mathcal{R}}
\newcommand{\bun}{\mathbf{1}}
\newcommand{\vph}{\varphi}
\newcommand{\vth}{\vartheta}
\newcommand{\vu}{{\vec{u}}}
\newcommand{\pp}{\phantom{+}}
\newcommand{\dd}{\partial}
\newcommand{\sd}{\mathcal{D}}
\newcommand{\oh}{\tfrac{1}{2}}

\renewcommand{\PaperNumber}{030}

\FirstPageHeading

\ShortArticleName{The Supersymmetric Burgers and Boussinesq
Equations}

\ArticleName{Supersymmetric Representations \\ and Integrable Fermionic Extensions\\
   of the Burgers and Boussinesq Equations}

\Author{Arthemy V. KISELEV~$^{\dag\ddag}$ and Thomas WOLF~$^\S$}

\AuthorNameForHeading{A.V.~Kiselev and Th.~Wolf}

\Address{$^\dag$~Department of Higher Mathematics, Ivanovo State
Power University,\\
$\phantom{^\dag}$~34 Rabfakovskaya Str., Ivanovo, 153003 Russia}

\Address{$^\ddag$~Department of Physics, Middle East Technical
University, 06531 Ankara, Turkey}

\EmailD{\href{mailto:arthemy@newton.physics.metu.edu.tr}{arthemy@newton.physics.metu.edu.tr}}

\Address{$^\S$~Department of Mathematics, Brock University, 500 Glenridge Ave.,\\
$\phantom{^\S}$~St.~Catharines, Ontario, Canada L2S~3A1}

\EmailD{\href{mailto:twolf@brocku.ca}{twolf@brocku.ca}}

\ArticleDates{Received November 26, 2005,
 in f\/inal form February 25, 2006;
Published online February 28, 2006}

\Abstract{We construct new integrable coupled systems of $N=1$
supersymmetric equations and present integrable fermionic
extensions of the Burgers and Boussinesq equations. Existence of
inf\/initely many higher symmetries is demonstrated by the
presence of recursion operators. Various algebraic methods are
applied to the analysis of symmetries, conservation laws,
recursion operators, and Hamiltonian structures. A fermionic
extension of the Burgers equation is related with the Burgers
f\/lows on associative algebras. A Gardner's deformation is found
for the bosonic super-f\/ield dispersionless Boussinesq equation,
and unusual properties of a recursion operator for its Hamiltonian
symmetries are described. Also, we construct a
three-parametric supersymmetric system that incorporates the
Boussinesq equation with dispersion and dissipation but never
retracts to it for any values of the parameters.}


\Classification{35Q53; 37K05; 37K10; 37K35; 58A50; 81T40}

\section{Introduction}
In this paper we construct new integrable coupled boson$+$fermion
systems of $N=1$ supersymmetric equations and present fermionic
extensions for the Burgers and Boussinesq equations. The
integrability of new systems is established by f\/inding weakly
non-local~\cite{Bilge, Novikov, Sergyeyev} recursion operators for
their symmetry algebras or by describing Gardner's
deformations~\cite{Gardner, PamukKale, KuperIrish}. We f\/ind
three integrable $N=1$ supersymmetric analogues of the KdV
equation. Also, we relate a fermionic extension of the Burgers
equation with the Burgers equations on associative
algeb\-ras~\cite{SokolovAssociative}. We apply algebraic
methods~\cite{JKKersten} to the study of geometry of
supersymmetric PDE, and we use the \textsc{SsTools}
package~\cite{CompPhysCommun} for the computer algebra system
\textsc{Reduce} in practical computations.

First we deal with generalizations of the Burgers equation, which
describes the dissipative nonlinear evolution of rarif\/ied gas.
We consider the bosonic super-f\/ield version of the Burgers
equation and construct two inf\/inite sequences of its symmetries;
the sequences correspond to even and odd `times' along the
f\/lows. Next, we analyze a family of fermionic extensions for the
Burgers equation itself. We show that if the coupling is zero,
then the system at hand reduces to the Burgers super-equation
w.r.t.\ a new f\/ield that unites the initial fermionic and
bosonic components. Otherwise, a new Grassmann independent
variable is introduced and the fermionic extensions are
transformed to the Burgers equation on an associative algebra~(see
also~\cite{SokolovAssociative}). Also, we observe that an $N=2$
generalization of the Burgers equation appears as a symmetry
f\/low for the Laberge--Mathieu's $N=2$ supersymmetric SKdV${}_4$
equation~\cite{Laberge}. The diagonal reduction
$\theta^1=\theta^2$ of this Burgers system is the $N=1$
super-f\/ield Burgers equation.

Further, we consider the systems related with the Boussinesq
equation, which describes the propagation of waves in a weakly
nonlinear and weakly dispersive liquid. We construct a Gardner's
deformation~\cite{Gardner, KuperIrish} of the dispersionless
bosonic super-f\/ield Boussinesq equation. Thus we recursively
def\/ine its Hamiltonians, which are further transmitted to the
boson$+$fermion representation of the system at hand.
Independently, we construct an unusual weakly
non-local~\cite{Bilge, Sergyeyev} recursion operator for this
system: its dif\/ferential order is~$1$, although it proliferates
the symmetries of constant order~$2$. These symmetries are
Hamiltonian w.r.t.\ the previously found chains of functionals,
and the `times' along the f\/lows are even and odd variables,
respectively. Finally, we extend a super-f\/ield representation
for the `full' Boussinesq equation with dispersion and dissipation
to a family of coupled boson$+$fermion evolutionary super-systems
that contain the Boussinesq equation but do not retract to it at
any values of the parameters.

The paper is organized as follows. In the introductory part below,
we describe three analogues of the $N=1$ supersymmetric KdV
equation~\cite{ManinRadul, Mathieu} and construct weakly non-local
recursion operators for their symmetry algebras, see
Example~\ref{SecKdV} on p.~\pageref{SecKdV} and
Example~\ref{ShadowTrue} in Section~\ref{SecRec}. Also, we discuss
the properties of a non-local Gardner's deformation~\cite{Andrea,
Gardner} for the $N=1$ supersymmetric KdV equation itself. In
Section~\ref{SecRec} we recall two schemes for generating
inf\/inite sequences of higher symmetries of the evolutionary
super-systems which are contained in the experimental
database~\cite{SUSY} and which are the objects studied in this
paper. In Section~\ref{SecBurg} we investigate super-f\/ield
representations and fermionic extensions of the Burgers equation.
Then in Section~\ref{SecBous} we discuss a Gardner's deformation
of the dispersionless Boussinesq equation, and we construct a
parametric family of super-systems that incorporate the Boussinesq
equation with dispersion and dissipation.

\subsection[The classification problem]{The classif\/ication problem}
The motivating idea of this research is the problem of a complete
description of $N=1$ supersymmetric nonlinear scaling-invariant
evolutionary equations $\{f_t=\phi^f, \, b_t=\phi^b\}$ that admit
inf\/initely many local symmetries proliferated by recursion
operators; here $b(x,t,\theta)$ is the bosonic super-f\/ield and
$f(x,t,\theta)$ is the fermionic super-f\/ield. We denote by
$\theta$ the super-variable and we put $\sd\equiv
D_\theta+\theta\,D_x$ such that $\sd^2=D_x$ and
$[\sd{},\sd{}]=2D_x$;
here $D_\theta$ and $D_x$ are the 
derivatives w.r.t.\ $\theta$ and $x$, respectively (fortunately,
the derivative $D_\theta$ is met very seldom in the text, hence no
confusion with the operator~$\sd$ occurs). The following axioms
suggested by V.V.~So\-ko\-lov and A.S.~So\-rin were postulated:
\begin{enumerate}\label{pAxioms}\vspace{-2mm}
\itemsep=0pt \item Each equation admits at least one higher
symmetry $\binom{f_s}{b_s}$. 
\item All equations are translation
invariant and do not depend on the time $t$ explicitly. 
\item
\label{ax3} None of the evolution equations involves only one
f\/ield and hence none of the r.h.s.\ vanishes. 
\item  \label{ax4}
At least one of the right-hand sides in either the evolution
equation or its symmetry is nonlinear. 
\item \label{AxN0N1} At
least one equation in a system or at least one component of its
symmetry contains a~fermionic f\/ield or the
super-derivative~$\sd{}$. 
\item The time $t$ and the parameters
$s$ along the integral trajectories of the symmetry f\/ields are
even variables (that is, the parities of $b$ and $b_t$ or $b_s$
coincide, as well as the parities of~$f$,~$f_t$, and $f_s$). 
\item
\label{AxScale} The equations are scaling invariant: their
right-hand sides are dif\/ferential polynomials homogeneous
w.r.t.\ a set of (half-)integer weights $[\theta]\equiv-\oh$,
$[x]\equiv-1$, $[t]<0$, $[f]$, $[b]>0$; we also assume that the
negative weight $[s]$ is (half-)integer.\vspace{-2mm}
\end{enumerate}

Axiom~8 for $N\geq2$ supersymmetric equations that satisfy the
above axioms is further described on p.~\pageref{ax8}. If a
particular equation under study admits symmetry f\/lows with an
\emph{odd} `time', then we use the notation $\bar{s}$ and the
parities of $f_{\bar{s}}$, $b_{\bar{s}}$ are opposite to the
parities of $f$ and~$b$, respectively.

\begin{remark}
From Axiom~\ref{AxN0N1} it follows that the admissible systems are
either fermionic extensions of bosonic systems, or they are
$N\geq1$ supersymmetric super-f\/ield equations and the
derivations~$\sd{}$ are present explicitly. For instance, the
superKdV equation~\eqref{sKdV} w.r.t.\ a fermionic super-f\/ield
$f(x,t,\theta)$ is $N=1$ supersymmetric, see~\cite{ManinRadul,
Mathieu}. On the other hand, an $N=0$ two-component fermionic
extension of the Burgers equation that can not be represented as
 a~scalar $N=1$ equation is obtained in Section~\ref{SecBurgFamily},
see equation~\eqref{SuperBurg} on p.~\pageref{SuperBurg}.
\end{remark}

The f\/irst version of the \textsc{Reduce} package
\textsc{SsTools} by T.~Wolf and W.~Neun for super-calculus allowed
to do symmetry investigations of supersymmetric equations. It was
used for f\/inding scalar fermionic and bosonic super-f\/ield and
$N=1$ supersymmetric equations, and for description of coupled
fermion$+$fermion, fermion$+$boson, and boson$+$boson evolutionary
systems that satisfy the above axioms; a number of $N=2$ scalar
evolution PDE were also found. The bounds $0<[f]$, $[b]\leq5$ and
$0>[t]>[s]\geq-5$ were used. The experimental database~\cite{SUSY}
contains $1830$ equations (the duplication of PDE that appeared
owing to possible non-uniqueness of the weights is now eliminated)
and their $4153$ symmetries (plus the translations along $x$ and
$t$, and plus the scalings whose number is in fact inf\/inite).

\begin{remark}\label{AxTooRestrictive}
Axiom~\ref{AxScale} together with Axiom~\ref{ax3} are very
restrictive. Indeed, a class of integrable systems that do not
satisfy these two assumptions is provided by the Gardner's
deformations, see~\cite{Gardner} and~\cite{PamukKale, KuperIrish}.
The extended $N=1$ superKdV${}_\varepsilon$ equation~\eqref{sKdVe}
and equation~\eqref{HydroE} on p.~\pageref{HydroE} give two
examples; many other completely integrable extended systems are
found in \textit{loc.\ cit}. We emphasize that no scaling
invariance can be recognized for the Gardner's extended systems if
non-zero values of the deformation parameters~$\varepsilon$ are
f\/ixed. The scaling invariance is restored if the weights of the
parameters are assumed non-zero: one has $[\varepsilon]=-1$ for
equation~\eqref{sKdVe} and $[\varepsilon]=-3$ for
system~\eqref{HydroE}. Hence we conclude that a classif\/ication
(see~\cite{Tsuchida} and references therein) of the `symmetry
integrable' homogeneous evolution equations may be incomplete,
providing only zero order terms of the deformations
in~$\varepsilon$.

Also, we note that the symmetry integrability approach, which was
used to f\/ill in the database~\cite{SUSY}, has revealed a number
of systems whose integrability in any sense remains an open
problem. For example, supersymmetric equation~\eqref{BousEmbed}
admits four symmetries, but this knowledge can hardly contribute
to constructing a solution of the system at hand.

Equations~\eqref{sKdVe} and~\eqref{BousEmbed} demonstrate that the
database~\cite{SUSY} with supersymmetric and coupled
boson$+$fermion systems is not exhaustive and, simultaneously, it
may contain equations whose complete or Lax integrability is
uncertain.
\end{remark}

\subsection[The $N=1$ superKdV equation and its extensions]{The $\boldsymbol{N=1}$ superKdV equation and its extensions}
The classical integrable supersymmetric evolutionary systems as
well as their generalizations and reductions are present in the
database.

\begin{example}[$\boldsymbol{N=1}$ superKdV~\cite{ManinRadul,Mathieu}]\label{SecKdV}
Let $N=1$ and let the weight of the bosonic
super-f\/ield~$b(x,t,\theta)$ be $[b]=1$; further, let $[t]=-3$.
Now we scan the cell which is assigned to these weights and which
is f\/illed in by the runs of \textsc{SsTools}. Then we get the
potential $N=1$ superKdV equation\footnote{Throughout this text,
the operator~$\sd{}$ acts on the succeeding super-f\/ield unless
stated otherwise explicitly.}
\begin{gather}\label{psKdV}
b_t=b_{xxx}+3\sd{(b_x\sd{b})}.
\end{gather}
Indeed, put $f(x,t;\theta)=\sd{b}$; then $f$ is the fermionic
super-f\/ield of weight $[f]=\tfrac{3}{2}$ satisfying the superKdV
equation~(\cite{ManinRadul, Mathieu}, see also~\cite{N1KdVDemo}
and~\cite[Ch.\,6]{JKKersten})
\begin{gather}\label{sKdV}
f_t=f_{xxx}+3(f\sd{f})_x.
\end{gather}

The potential super-f\/ield equation~\eqref{psKdV} admits two
conserved densities $\rho_1=b$, $\rho_2=\tfrac{1}{2}b^2$, which
are nonlocal w.r.t.\ equation~\eqref{sKdV}; also, potential
equation~\eqref{psKdV} inherits inf\/initely many fermionic
conservation laws from superKdV equation~\eqref{sKdV}. The local
conserved densities for~\eqref{sKdV} are obtained by inverting the
Miura transformation (see~\cite{KuperIrish, Gardner})
$f=\chi+\varepsilon\chi_x-\varepsilon^2\chi\sd{\chi}$ from the
Gardner's extended $N=1$ superKdV${}_\varepsilon$
equation~\cite{Mathieu, MathieuOpen}
\begin{gather}\label{sKdVe}
\chi_t = \chi_{xxx} + 3\bigl(\chi\sd{\chi}\bigr)_x
   - \tfrac{1}{2}\varepsilon^2\sd\left[\bigl(\sd{\chi}\bigr)^3\right]
 - \tfrac{3}{2}\varepsilon^2\left[\chi\bigl(\sd{\chi}\bigr)^2\right]_x.
\end{gather}
Recently in~\cite{Andrea} it has been observed that the bosonic
non-local conserved densities for~\eqref{psKdV} and~\eqref{sKdV}
are obtained by using an appropriate change of coordinates in the
potential superKdV${}_\varepsilon$ equation. Namely, let us
introduce the bosonic super-f\/ield~$\phi(x,t,\theta;\varepsilon)$
such that $\chi=\sd{\phi}$. Note that the evolution equation
for~$\phi$ is not in the form of a conserved current,
\begin{gather}\label{psKdVe}
\phi_t = \phi_{xxx} + 3\sd\bigl(\sd\phi\cdot{\phi}_x\bigr)
   - \tfrac{1}{2}\varepsilon^2\phi_x^3
 - \tfrac{3}{2}\varepsilon^2\sd\bigl[\sd\phi\cdot{\phi}_x^2\bigr].
\end{gather}
Nevertheless, put
$\psi(x,t,\theta;\varepsilon)=\exp(\varepsilon\phi)$. Then
$\psi(x,t,\theta;\varepsilon)$ satisf\/ies the equation
\begin{gather}\tag{\ref{psKdVe}${}'$}\label{ExppsKdVe}
\psi_t = \sd\left[\sd{\psi_{xx}}
   - \frac{3\psi_x\sd\psi_x}{\psi}
   + \frac{1}{\varepsilon}\cdot\frac{3\sd\psi\,\psi_x}{\psi}
\right].
\end{gather}
Potential extended superKdV${}_\varepsilon$
equation~\eqref{ExppsKdVe} was implicitly described
in~\cite{Andrea}. We note that equation~\eqref{ExppsKdVe} is
\emph{singular} in the deformation parameter~$\varepsilon$. This
is rather unusual with respect to the common practice
(see~\cite{KuperIrish} and references therein), although singular
transformations did appear~\cite{KuperDiscrete} in the context of
$\varepsilon$-parametric families of integrable equations.

The Taylor coef\/f\/icients $\psi_n$ of the standard decomposition
$\psi=\sum\limits_{n=1}^{+\infty}\varepsilon^n\cdot\psi_n[f]$ are
the required~\cite{MathieuOpen} nonlocal conserved densities
for~\eqref{sKdV}. The transformation $f=f[\psi]$ def\/ines the
recurrence relation upon them, and the initial condition is
$\psi_1=b$. Hence we obtain $\rho_1$, $\rho_2$,
$\rho_3=\tfrac{1}{6}b^3+\sd^{-1}(f\sd{f})$, etc. The new densities
$\rho_n$, $n\geq3$ are still non-local w.r.t.\ potential superKdV
equation~\eqref{psKdV}.
\end{example}

\subsection[Analogues of the $N=1$ super KdV equation]{Analogues of the $\boldsymbol{N=1}$ super KdV equation}

Now we describe three analogues of superKdV equation~\eqref{sKdV}.
These fermionic super-f\/ield equations are homogeneous w.r.t.\
the same weights as~\eqref{sKdV}, and they admit inf\/initely many
higher symmetries proliferated by weakly non-local~\cite{Bilge,
Sergyeyev} recursion operators. To this end, let the weights
$[f]=\tfrac{3}{2}$ and $[t]=-3$ be f\/ixed. Then we obtain four
evolutionary supersymmetric equations (namely,~\eqref{sKdV}
and~(\ref{eqsNaumov}a,b,c)) that admit higher symmetries
$f_s=\phi$ under the assumption $[s]\geq-5$. The superKdV
equation, see~\eqref{sKdV}, is the f\/irst in this list. We also
get the equation
\begin{subequations}\label{eqsNaumov}
\begin{gather}
f_t=f_{xxx}+f_x\sd{f}. \label{eq2}
\end{gather}
The recursion operator for equation~\eqref{eq2} is constructed in
Example~\ref{ShadowTrue} on p.~\pageref{ShadowTrue}. Third, we
obtain the two-parametric dispersionless analogue of
equation~\eqref{sKdV}:
\begin{gather}
f_t=\alpha\,f\sd{f_x} + \beta\,f_x \sd{f},\qquad
\alpha,\beta=\const. \label{eq3}
\end{gather}
\end{subequations}
A computation by Yu.~Naumov (Ivanovo State Power University) with
\textsc{SsTools} demonstrates that equation~\eqref{eq3} admits the
weakly non-local recursion operator
\begin{gather}\label{RecEq3}
R=\alpha f\sd{f}\cdot\sd{} + \alpha f f_x -\alpha f
\sd{f_x}\cdot\sd^{-1} -\beta f_x\sd{f}\cdot\sd^{-1} + \beta
f_x\cdot\sd^{-1}\circ(f\sd + \sd{f})
\end{gather}
and two inf\/inite sequences of symmetries that start from the
translations $f_x$ and $f_t$.

The fourth equation for the set of weights $[f]=\tfrac{3}{2}$,
$[t]=-3$ is
\begin{gather}\tag{\ref{eqsNaumov}c}\label{eq4}
f_t=\sd{(f_x f)}.
\end{gather}
It admits the recursion $R=f\sd{} -f_x\sd^{-1}$, and also it has
an inf\/inite sequence of symmetries that starts from the odd
weight~$[\bar{s}]=-8\oh$.

\begin{remark}
The symbols of the evolutionary supersymmetric equations that
possess inf\/initely many symmetries are not necessarily constant.
For example, equation~\eqref{eq4} can not be transformed to an
equation $g_t=g_{xxx}+\cdots$ by a dif\/ferential substitution
$f=f[g]$. The proof is by \textit{reductio ad absurdum}.
\end{remark}

\subsection{Remarks}
In this paper, we investigate the geometric properties of the
boson$+$fermion systems under the additional assumption $[f]=[b]$
(for the primary sets of weights if they are multiply def\/ined).
From Axioms~\ref{ax3} and~\ref{ax4} on p.~\pageref{pAxioms} it
follows that the triangular systems are regarded as trivial and
therefore their properties are not analyzed. We emphasize that,
generally, we can not introduce a new anticommuting
variable~$\varTheta$ and then unite the two super-f\/ields $f$,
$b$ to the fermionic super-f\/ield $\phi=f+\varTheta b$ such that
$[\phi]=[f]=[b]-\oh$ or to the bosonic super-f\/ield
$\beta=b+\varTheta f$ such that $[\beta]=[b]=[f]-\oh$ and such
that a scalar equation w.r.t.\ $\phi$ or $\beta$ holds (there is
no contradiction with the diagonality assumption because the
weights may not be uniquely def\/ined).

We do not expose now the complete list of supersymmetric
bo\-son$+$fer\-mi\-on systems that satisfy the Axioms on
p.~\pageref{pAxioms} and such that the weights $[f]=[b]$ coincide.
In fact, the symmetries for a major part of these equations are
proliferated by the recurrence relations (see
p.~\pageref{DefRecRelation}); other equations that admit true
recursions seem less physically important than the three variants
of the Burgers and the Boussinesq equations we analyze.

Yet it is worthy to note some remarkable features of the f\/ive
systems such that the weight $[t]=-\oh$ of the time $t$ is half
the weight of the spatial variable~$x$ (that is, the equations
precede the translation invariance). It turned out that these
f\/ive equations exhibit practically the whole variety of
properties that superPDE of mathematical physics possess. Let us
brief\/ly summarize these features.

Three of the f\/ive evolutionary systems are given through
\begin{gather}\label{Quad}
f_t=-\alpha fb,\qquad b_t=b^2+\sd{f},\qquad \alpha=1,2,4.
\end{gather}
The equations dif\/fer by the values $\alpha=1$, $2$, $4$ of the
coef\/f\/icient and demonstrate dif\/ferent geometrical
properties. The geometry of the $\alpha=2$ system is quite
extensive: this system admits a continuous sequence of symmetries
for all (half-)integer weights $[s]\leq-\oh$, a sequence of
symmetries such that the parities of the dependent variables are
opposite to the parities of their f\/lows, four local recursions
(one is nilpotent), and three local super-recursions. The equation
for $\alpha=1$ admits fewer structures, and the case $\alpha=4$
for equation~\eqref{Quad} is rather poor.

Another equation
\begin{gather}\label{DoubleLayer}
f_t=\sd{b}+fb,\qquad b_t=\sd{f}
\end{gather}
admits local symmetries for all (half-)integer weights
$[s]\leq-\oh$. Equation~\eqref{DoubleLayer} requires introduction
of two layers of nonlocalities assigned to (non)local conservation
laws. Four nonlocal recursion operators with nonlocal
coef\/f\/icients are then constructed for
equation~\eqref{DoubleLayer}. The pro\-per\-ties of systems
(\ref{Quad}), (\ref{DoubleLayer}) are considered in details in the
succeeding paper~\cite{Dubna}.

The f\/ifth system we mention is a super-f\/ield representation of
the Burgers equation, see~\eqref{BurgSystem}; this system is
$C$-integrable by using the Cole--Hopf substitution. We
investigate its properties in Section~\ref{SecBurg}.

\section{Recursion operators and recurrence relations}\label{SecRec}

In this section we describe two principally dif\/ferent mechanisms
for proliferation of symmetries of a PDE.

\subsection[Differential recursion operators]{Dif\/ferential recursion operators}
We consider the (nonlocal) dif\/ferential recursion operators
f\/irst.
The standard approach \cite{N1KdVDemo, JKKersten} to recursion
operators is regarding them as symmetries of the linearized
equations. The essence of the method is the following. The
`phantom variables' (the Cartan forms) that satisfy the linearized
equation are assigned to all the dependent variables in an
equation~$\cE$; one may think that the internal structure of the
symmetries is discarded and the (nonlocal) phantom variables
imitate the (resp., nonlocal components of) symmetries for~$\cE$.
In what follows, the capital letters $F$, $B$, etc.\ denote the
variables associated with the f\/ields $f$, $b$, respectively.
Then any image $\cR=R(\varphi)$ of a linear operator $R$ that maps
symmetries $\varphi=\{f_s=F$, $b_s=B\}$ of $\cE$ to symmetries
again is linear w.r.t.\ the right-hand sides $F$, $B$. One easily
checks that $\cR$ is then the r.h.s.\ of a symmetry f\/low
\[
\frac{d}{ds_R}\binom{F}{B}=\cR
\]
for the linearized equation $\Lin(\cE)$. If the initial equation
$\cE$ is evolutionary, then the phantom variables satisfy the
well-known relations
\[
F_{[t,s]}\doteq0,\qquad B_{[t,s]}\doteq0
\]
that hold by virtue ($\doteq$) of the equations $\cE$ and
$\Lin(\cE)$. The method is reproduced literally in presence of
nonlocalities~$w$ whose f\/lows~$W$ are described by the
corresponding components of nonlocal symmetries
$\hat{\vph}=(f_s=F$, $b_s=B$, $w_s=W)$. The recursion operators
are then def\/ined by the triples $\hat{\cR}=(F_{s_R}$, $B_{s_R}$,
$W_{s_R})$ and generate sequences of nonlocal symmetries.
See~\cite{N1KdVDemo, JKKersten} for many examples.

We f\/inally recall that not each symmetry $\vph$ can be extended
to a nonlocal f\/low $\hat{\vph}$ if the set $\{w\}$ of
nonlocalities is already def\/ined and, analogously, not all the
pairs $\cR=(F_{s_R}$, $B_{s_R})$ generate a~true
recursion~$\hat{\cR}$. The pairs $\cR$ are therefore called
\emph{shadows}~\cite{JKKersten} of nonlocal recursion operators.
The shadows are usually suf\/f\/icient for standard purposes if
they describe the operators that map the local components of the
f\/lows and whose coef\/f\/icients are also local. Hence in what
follows we always set $W_{s_R}=0$ (that is, we do not f\/ind the
f\/lows $W_{s_R}$ that commute with the evolution~$W_t$ determined
by the original system $\cE$ and dif\/ferential substitutions
for~$w$). Also, we describe the Cartan forms $\cR$ rather than the
dif\/ferential operators~$R$, and we use the term `recursions'
instead of the rigorous `shadows of the generating Cartan forms
for nonlocal recursion operators.'

\begin{remark}\label{RemSergyeyev}
The recursion operators considered in this paper are weakly
non-local, see~\cite{Bilge, Novikov}. Hence one can readily prove
the locality of symmetry sequences generated by these recursions
by using a supersymmetric version of the results
in~\cite{Sergyeyev}. In the sequel, we use another method for the
proof of locality. Namely, in Section~\ref{BosonicDBous} we
construct a Gardner's integrable deformation~\cite{PamukKale,
KuperIrish, Gardner} and thus we obtain local Hamiltonian
functionals, whence we deduce the locality of the corresponding
symmetry f\/lows.
\end{remark}

We f\/inally note that a similar method of `phantom variables' is
applied for f\/inding Hamiltonian and symplectic structures for
PDE, see~\cite{Lstar}. The (nonlocal) Hamiltonian structures for
supersymmetrizations~\eqref{BousParam}, \eqref{BousEmbed} are not
extensively studied in this paper. The \textsc{SsTools} packa\-ge
is applicable for this investigation \emph{as is}, since the
theory is now transformed to standard algorithms of symmetry
analysis.

\begin{example}\label{ShadowTrue}
Consider analogue~\eqref{eq2} of the superKdV
equation~\eqref{sKdV}. We introduce the bosonic nonlocality $v$ of
weight $[v]=1$ such that $\sd{v}=f$ and the fermionic nonlocality
$w$ such that $[w]=\tfrac{7}{2}$ and $\sd{w}=(\sd{f})^2$. In this
setting, we obtain the recursion
\[
\cR_{[2]} = \bigl( \sd{f}\cdot F + 3F_{xx} + f_x\cdot V
+\tfrac{1}{2}W \bigr).
\]
The above solution generates the sequence $f_x\mapsto
f_t\mapsto\cdots$ of symmetries for equation~\eqref{eq2}. The
sequence starts with the translation along~$x$ and next contains
the equation itself.

Next, we consider equation~\eqref{eq3} and introduce two bosonic
nonlocal variables $v$ and $w$ such that $\sd{v}=f$ and
$\sd{w}=f\sd{f}$. Then we obtain the nonlocal recursion
\[
\cR=\alpha\cdot\left(\sd{f_x}f\,V-f\sd{f}\sd{F}+f_xfF \right) +
\beta\cdot\left(f\sd{f}\,V-f_xW\right).
\]
The corresponding operator~$R$ is present in~\eqref{RecEq3} on
p.~\pageref{RecEq3}.

Finally, equation~\eqref{eq4} is obviously a continuity relation.
Therefore, we let $v$ be the bosonic variable such that
$\sd{v}=f$; hence we obtain the recursion
$F_{s_R}=f\,\sd{F}-f_x\,V$.
\end{example}

\begin{remark}
The systems that admit several scaling symmetries and hence are
homogeneous w.r.t.\ dif\/ferent weights allow to apply the breadth
search method for recursions, which is the following. Let a
recursion of weight $[s_R]$ w.r.t.\ a particular set of weights
for the super-f\/ields~$f$,~$b$ and the time $t$ be known. Now,
recalculate its weight $[s'_R]$ w.r.t.\ another set and then
f\/ind all recursion operators of weight~$[s'_R]$. The list of
solutions will incorporate the known recursion and, possibly,
other operators. Generally, their weights will be dif\/ferent from
the weight of the original recursion w.r.t.\ the initial set.
Hence we repeat the reasonings for each new operator and thus
select the weights $[s_R]$ such that nontrivial recursions exist.
This method is a serious instrument for the control of
calculations and elimination of errors. We used it while testing
the second version of the \textsc{SsTools}
package~\cite{CompPhysCommun}.
\end{remark}

The second version of \textsc{SsTools} allows to reduce the search
for nonlocal recursion operators to solving large overdetermined
systems of nonlinear algebraic equations for the undetermined
coef\/f\/icients which are present in the weight-homogeneous
ansatz for $F_{s_R}$, $B_{s_R}$. The algebraic systems are then
solved by using the program \textsc{Crack}~\cite{WolfCrack}. The
nonlocal variables, which are assigned to conservation laws if
$N\leq1$, were
also obtained by \textsc{SsTools} straightforwardly using the 
weight homogeneity assumptions.

\begin{remark}
The weights of the nonlocal variables constructed by using
conserved currents for PDE are def\/ined by obvious rules.
Clearly, if the weight for a bosonic nonlocality is zero, then
further assumptions about the maximal power of this variable in
any ansatz should be made. Within this research we observed that
the weights of the new super-f\/ields necessary for constructing
the recursions are never negative.
\end{remark}

\subsection{Recurrence relations}
Consider an evolution equation $\cE=\{u_t=\phi\}$ and let there be
a dif\/ferential function $q[f,b]$ of weight zero w.r.t.\ an
admissible set of weights for~$\cE$. Suppose further that the
f\/low $u_{s^n}=q^n\cdot\phi$ is a symmetry of $\cE$ for any
$n\in\BBN$; in a typical situation all the f\/lows
$\vph_n=u_{s^n}$ commute with each other. Then, instead of an
inf\/inite sequence $\vph_n$ we have just one right-hand side
$u_s=Q(q)\cdot\phi$ of f\/ixed dif\/ferential order and
weight~$[\phi]$; here $Q$ is an arbitrary analytic function. In
this case, we say that the sequence of the Taylor monomials
$\vph_n$ is generated by a \emph{recurrence
relation}.\label{DefRecRelation}

We note that the multiplication by $q$ can be a zero-order
recursion operator $R=q$ for the whole symmetry algebra $\sym\cE$,
otherwise the recurrence relation $\vph_{n+1}=q\cdot\vph_n$
generates the symmetries of $\cE$ for the f\/ixed `seed'
f\/lows~$\vph_0$. The systems that admit recurrence relations for
their symmetries can possess dif\/ferential recursion operators as
well.

\begin{example}  
Consider the family of supersymmetric systems
\begin{gather}\label{Hospital}
f_t=b\sd{b}+f\sd{f},\qquad b_t=\alpha f\sd{b},  
\end{gather}
here $\alpha\in\BBR$ is arbitrary. We see that
equation~\eqref{Hospital} is homogeneous w.r.t.\ the weights
$[f]=[b]=\oh$, $[t]=-1$. The multiplication of the r.h.s.\
in~\eqref{Hospital} by $b$ def\/ines a recurrence relation for
inf\/initely many symmetries. Indeed, the f\/lows
\[
f_t=b\,Q(b)\cdot\sd{b}+f\,Q(b)\cdot\sd{f},\qquad b_t=\alpha
f\,Q(b)\cdot\sd{b}
\]
commute for all $Q$s and any constant~$\alpha$.
Nevertheless, the operator $\left(\begin{smallmatrix}b&0\\
0&b\end{smallmatrix}\right)$ is not a recursion for
equation~\eqref{Hospital} because it does not map an
\emph{arbitrary} symmetry to a symmetry.

Further, let $\alpha=1$. The system
\begin{gather}\tag{\ref{Hospital}a}\label{Diag007Hospital}
f_t=b\sd{b}+f\sd{f},\qquad b_t= f\sd{b}
\end{gather}
admits the local zero-order recursions
\begin{gather*}
\cR_{[1\frac{1}{2}]}^1=
   \binom{\sd{b}bB+\sd{b}fF-\sd{f}fB}{\sd{b}fB},\qquad
\cR_{[2]}^2=
   \binom{b_xbF+f_xbB}{b_xbB},\\
\cR_{[3]}^3=
   \binom{\sd{b}b_xbB+\sd{b}b_xfF-\sd{b}f_xfB-\sd{f}b_xfB}%
         {\sd{b}\,b_xfB}.
\end{gather*}
An inf\/inite number of local recursion operators for
equation~\eqref{Diag007Hospital} is obtained by multiplication of
$\cR^1$ by $b^n$, $n\in\BBN$. The recursions $\cR^1$ and $\cR^3$
are {nilpotent}: $\left(\cR_{[1\frac{1}{2}]}^1\right)^4=0=
\left(\cR_{[3]}^3\right)^4$.

If $\alpha=-1$, then equation~\eqref{Hospital} also admits
inf\/initely many symmetries that do not originate from any
recurrence relation because their dif\/ferential orders grow.
\end{example}

The recurrence relation $\vph_{n+1}(\vph_n$,\ $q$,\ $n)$ can
depend explicitly on the subscript~$n$; then the generators of
commuting f\/lows contain the free functional parameters $Q(q)$,\
$Q'(q)$, etc.

\begin{example}
The f\/lows
\begin{gather}\label{CommuteWithPrime}
\binom{f_s(Q)}{b_s(Q)} =
\binom{\alpha f_xQ(b) + \gamma b_xfQ'(b) + \delta fb^2Q'(b)}%
   {\alpha b_xQ(b) + \beta f_xfQ'(b)}
\end{gather}
commute for arbitrary functions $Q(b)$ and constants $\alpha$,\
$\beta$,\ $\gamma$,\ $\delta\in\BBR$. Indeed, for any $Q(b)$
and~$S(b)$ we have
\[
\left\{
\binom{f_\tau(Q)}{b_\tau(Q)},\binom{f_\sigma(S)}{b_\sigma(S)}
\right\}=0.
\]
The f\/lows def\/ined in~\eqref{CommuteWithPrime} are also
translation and scaling invariant.
\end{example}

In the sequel, we investigate the systems that are located on the
diagonal $[f]=[b]$ and admit inf\/inite sequences of (commuting)
symmetries generated by recursion operators; we also analyze
generalizations of these equations and properties of the new
systems. The supersymmetric equations whose symmetries are
proliferated by the recurrence relations are {not} discussed in
this paper.


\section{The Burgers equation}\label{SecBurg}
In this section we investigate three systems related with the
Burgers equation. We consider an $N=1$ super-f\/ield
representation of the Burgers equation and analyze its symmetry
properties, we relate a fermionic extension of the Burgers
equation with the Burgers equation on associative algebras, and we
indicate an $N=2$ scalar super-equation whose $N=1$ diagonal
reduction ($\theta^1=\theta^2$) is the bosonic super-f\/ield
Burgers equation again.

\subsection[Super-field representation for the Burgers equation]{Super-f\/ield representation for the Burgers equation}
Consider the system 
\begin{gather}\label{BurgSystem}
f_t=\sd{b},\qquad b_t=b^2+\sd{f}.
\end{gather}
There is a unique set of weights $[f]=[b]=\oh$, $[t]=-\oh$,
$[x]=-1$ in this case. Hence we conclude that the above system
precedes the invariance w.r.t.\ the translation along~$x$.
Equation~\eqref{BurgSystem} admits the continuous
sequence~\eqref{SymBurg} of higher symmetries $f_s=\phi^f$,
$b_s=\phi^b$ at all
(half-)integer weights $[s]\leq-\oh$.    
Also, there is another continuous sequence~\eqref{BurgSSym} of
symmetries for equation~\eqref{BurgSystem} at all (half-)integer
weights $[\bar{s}]\leq-\oh$ of the odd `time'~$\bar{s}$.

System~\eqref{BurgSystem} is obviously reduced to the bosonic
super-f\/ield Burgers equation
\begin{gather}\label{BurgersInverse}
b_x=b_{tt}-2bb_t,\qquad b=b(x,t,\theta).
\end{gather}
We emphasize that the role of the independent coordinates $x$ and
$t$ is reversed w.r.t.\ the standard interpretation of $t$ as the
time and $x$ as the spatial variable. The Cole--Hopf substitution
$b=-u^{-1}u_t$ from the heat equation
\[
u_x=u_{tt}
\]
provides the solution for the bosonic component
of~\eqref{BurgSystem}.

Further, we introduce the bosonic nonlocality $w(x,t,\theta)$ of
weight $[w]=0$ by specifying its derivatives,
\[
\sd{w}=-f,\qquad  w_t=-b;
\]
the variable $w$ is a potential for both f\/ields $f$ and $b$. The
nonlocality satisf\/ies the potential Burgers equation
$w_x=w_{tt}+w_t^2$ such that the formula $w=\ln u$ gives the
solution; the relation $f=-\sd{w}$ determines the fermionic
component in system~\eqref{BurgSystem}.

Now we extend the set of dependent variables $f$, $b$, and $w$ by
the symmetry generators $F$, $B$, and $W$ that satisfy the
respective linearized relations upon the f\/lows of the initial
super-f\/ields. In this setting, we obtain the recursion
\begin{subequations}\label{RecBurgers}
\begin{gather}
\cR_{[1]}=\binom{F_x-\sd{f}\,F+f_x\,W}
                        {B_x-\sd{f}\,B+b_x\,W}
\end{gather}
of weight $[s_R]=-1$. The operator assigned to $\cR$ is
\begin{gather}
R=\begin{pmatrix}
D_x-\sd{f}+f_x\,\sd^{-1} & 0 \\
b_x\,\sd^{-1} & D_x-\sd{f}
\end{pmatrix}.
\end{gather}
\end{subequations}
The above recursion~$R$ is \emph{weakly non-local}~\cite{Bilge,
Novikov}, that is, each nonlocality $\sd^{-1}$ is preceded with a
(shadow \cite{JKKersten} of a nonlocal) symmetry $\vph_\alpha$ and
is followed by the gradient $\psi_\alpha$ of a conservation law:
$R=\text{local
part}+\sum_\alpha\vph_\alpha\cdot\sd^{-1}\circ\psi_\alpha$.
From~\cite{Bilge} it follows that this property is satisf\/ied by
all recursion operators which are constructed by using one layer
of the nonlocal variables assigned to conservation laws. The weak
non-locality of recursion operators is essentially used in the
proof of locality of the symmetry hierarchies they generate,
see~\cite{Sergyeyev} and Remark~\ref{RemSergyeyev} on
p.~\pageref{RemSergyeyev}.

Recursion~\eqref{RecBurgers} generates two sequences of higher
symmetries for system~\eqref{BurgSystem}:
\begin{gather}\label{SymBurg}
\binom{f_t}{b_t}\mapsto
\binom{\sd{b_x}-\sd{f}\sd{b}-f_xb}{\sd{f_x}-(\sd{f})^2-b^2\sd{f}+bb_x}
   \mapsto\cdots, \qquad
\binom{f_x}{b_x}\mapsto
\binom{f_{xx}-2\sd{f}f_x}{b_{xx}-2\sd{f}b_x}\mapsto\cdots.
\end{gather}
The same recursion~\eqref{RecBurgers} produces two inf\/inite
sequences of symmetries with the odd parameters~$\bar{s}$ for the
Burgers equation:
\begin{gather}\label{BurgSSym}
\binom{\sd{f}}{\sd{b}} \mapsto
\binom{\sd{f_x}-(\sd{f})^2-f_xf}{\sd{b_x}-\sd{f}\,\sd{b}-b_xf}
\mapsto\cdots, \qquad
\binom{f\sd{b}-b\,\sd{f}+b_x}{b\sd{b}-f\,\sd{f}+f_x-fb^2}
\mapsto\cdots.
\end{gather}

\begin{remark}\label{ProfitInverse}
System~\eqref{BurgSystem} is not a supersymmetric extension of
equation~\eqref{BurgersInverse}; it is a representation of the
bosonic super-f\/ield Burgers equation. The f\/lows
in~\eqref{SymBurg} become purely bosonic in the coordinates
$c=\sd{f}$, $b$. The standard recursion for the Burgers equation,
see~\eqref{RecScalarBurg} on p.~\pageref{RecScalarBurg}, acts
`across' the two sequences in~\eqref{SymBurg} and maps $(f_t,
b_t)\mapsto(f_x, b_x)$; again, we note that the independent
coordinates in~\eqref{BurgersInverse} are reversed
w.r.t.~\eqref{Burgers}. Surprisingly, the f\/low that succeeds the
translation along~$x$ in~\eqref{SymBurg} reappears
in~\eqref{BurgWeiss}.

However, from the above reasonings we prof\/it two sequences of
symmetries~\eqref{BurgSSym}, which are \emph{not} reduced to the
bosonic $(x,t)$-independent symmetries~\cite[\S~8.2]{Lychagin} of
the Burgers equation. We f\/inally recall that the Burgers
equation~\eqref{BurgersInverse} has inf\/initely many higher
symmetries that depend explicitly on the base coordinates $x$, $t$
but exceed the set of Axioms on p.~\pageref{pAxioms}.
\end{remark}

\subsection{Supersymmetric extension of the Burgers equation}
\label{SecBurgFamily}

The fermionic extension of the Burgers equation,
\begin{gather}\label{SuperBurg}
f_t=f_{xx}+(bf)_x,\qquad b_t=b_{xx}+bb_x+\alpha\,f_xf,\qquad
\alpha\in\BBR,
\end{gather}
is a unique extension of the Burgers equation that admits higher
symmetries and which was found by using \textsc{SsTools}. It must
be noted that equation~\eqref{SuperBurg} contains the unknowns
$f(x,t)$, $b(x,t)$, and it seems to have nothing to do with a
supersymmetry. The one-parametric family~\eqref{SuperBurg} admits
the symmetries $\binom{f_s}{b_s}$ at all negative integer weights
$[s]\leq-1$. We also note that system~\eqref{SuperBurg} is the
Burgers equation itself if the fermionic f\/ield~$f(x,t)$ is set
to zero.

In what follows, we distinguish the two cases: $\alpha=0$ and
$\alpha\in\BBR\setminus\{0\}$ in equation~\eqref{SuperBurg}. We
claim that the algebraic properties of corresponding
extensions~\eqref{SuperBurg} for the Burgers equation are then
dif\/ferent. Indeed, system~\eqref{SuperBurg} is triangular if the
coupling constant $\alpha$ equals zero, see Remark~\ref{TwoWorlds}
below.

Obviously, the weights $[b]=1$ are $[t]=-2$ are f\/ixed in
system~\eqref{SuperBurg}. If $\alpha=0$, then the weight $[f]$ of
the fermionic f\/ield is arbitrary since the f\/irst equation
in~\eqref{SuperBurg} is linear w.r.t.\ the derivatives of~$f$. If
$\alpha\neq0$, then the weights of the f\/ield $f(x,t)$ and of the
\emph{dimensional} real constant~$\alpha$ are related by the
formula $[f]=1-\tfrac{1}{2}[\alpha]$, see
Remark~\ref{AxTooRestrictive} on p.~\pageref{AxTooRestrictive} for
discussion. In particular, we have $[f]=1$ if $\alpha$ is a true
constant w.r.t.\ the scalings and hence its weight equals zero. We
conclude that the f\/ields $b$ and $f$ can not be coupled to the
\emph{super}-f\/ield $u=b+\vartheta f$ using a
super-variable~$\vartheta$ unless $[f]=\tfrac{3}{2}$ and an
additional relation unites the weights of $\vartheta$
and~$\alpha$. In the sequel, we consider both cases:
$[f]=\tfrac{3}{2}$ and~$[f]\neq\tfrac{3}{2}$.

The fermionic component in~\eqref{SuperBurg} is linear w.r.t.\ the
f\/ield~$f(x,t)$, and hence the superposition principle is valid
for it. The quantity
\[
\int_{-\infty}^{+\infty} f(x,t)\,\Id x=\const
\]
is an integral of motion for~\eqref{SuperBurg}. The fermionic
variable $w(x,t)$ is assigned to the conserved current
$D_t(f)=D_x(f_x+bf)$: we set $w_x=f$ such that $w\cdot w=0$ and
$[w]=0$.

\begin{remark}\label{TwoWorlds}
System~\eqref{SuperBurg} models an unusual physical phenomenon.
Assume that at any point $x\in\BBR$ there are two types of a
physical value, the bosonic f\/ield with density $b(x,t)$ and the
``invisible'' fermionic f\/ield with density $w(x,t)$, and let the
dynamics of these two f\/ields be described by
system~\eqref{SuperBurg} (or equation~\eqref{BurgWeiss} below).
Suppose that the initial numeric values $w(x,0)$ and $b(x,0)$ of
the fermionic and bosonic densities, respectively, coincide at
$t=0$. If $\alpha=0$, then densities will coincide for all $t>0$
and the corresponding integral quantities
$\int_{-\infty}^{+\infty} w(x,t)\,\Id x$ and
$\int_{-\infty}^{+\infty} b(x,t)\,\Id x$ will be conserved in
time.

If $\alpha\neq0$, then the feed-back is switched on
in~\eqref{SuperBurg}. A ripple in the fermionic space is the cause
for the bosonic integral $\int_{-\infty}^{+\infty} b(x,t)\,\Id x$
to change. Indeed, this quantity is no longer conserved unless
$w(x,t)=\const$ or, generally, unless the condition
$\int_{-\infty}^{+\infty}w_{xx}(x,t)w_x(x,t)$ $\Id x=\const$ holds
for all $t>0$. We see that the reaction of the fermionic component
on the bosonic f\/ield depends on the incline $w_x=f$ and
curvature~$w_{xx}=f_x$ but not on the density~$w(x,t)$.
\end{remark}

\subsubsection[Trivial coupling in~(\ref{SuperBurg}): $\alpha=0$]{Trivial coupling in~(\ref{SuperBurg}):
$\boldsymbol{\alpha=0}$}

Let us suppose that $\alpha=0$. Then from~\eqref{SuperBurg} we
obtain the system
\begin{gather}\label{BurgWeiss}
w_t=w_{xx}+bw_x,\qquad b_t=b_{xx}+bb_x.
\end{gather}
System~\eqref{BurgWeiss} appeared in~\cite{Svinolupov}
and~\cite{Weiss} in a dif\/ferent context: both f\/ields $w(x,t)$
and $b(x,t)$ were regarded as bosonic, and then B\"acklund
autotransformations~\cite{Weiss} and the linearizing
substitutions~\cite{Svinolupov} were constructed. In
Remark~\ref{ProfitInverse} we noted that the two-component bosonic
system~\eqref{BurgWeiss} originates from a f\/low
in~\eqref{SymBurg}. The f\/ield~$w(x,t)$ was recognized as
fermionic in~\cite{Hlavaty}, where the Painlev\'e properties
of~\eqref{BurgWeiss} and related systems were investigated. In
what follows, the f\/ield~$w(x,t)$ is a fermionic dependent
variable.

\begin{remark}
The bosonic component $b(x,t)$ in~\eqref{BurgWeiss} is linearized
by the Cole--Hopf substitution
\begin{gather}\label{ColeHopf}
b=2q^{-1}q_x,
\end{gather}
where the function $q(x,t)$ is a solution of the heat equation
$q_t=q_{xx}$. From~\eqref{ColeHopf} it follows that the Burgers
equation is the factor of the heat equation w.r.t.\ its scaling
symmetry. This scheme is of general nature and can be used for
constructing new equations from homogeneous systems. In this paper
we do not investigate the relation between the scaling invariance
of superPDE and their factorizations w.r.t.\ the scaling
symmetries, and we do not study the physical signif\/icance of
these projectivizations and of the new equations.
\end{remark}

{\samepage Now we reduce~\eqref{SuperBurg} for $\alpha=0$ to one
scalar super-equation w.r.t.\ a new f\/ield which is constructed
as follows. Let $\vartheta$ be the new independent super-variable
such that
\begin{itemize}\vspace{-2mm}\itemsep=0pt
\item its square vanishes, $\vartheta\cdot\vartheta=0$; \item the
variable $\vartheta$ anticommutes with the fermionic f\/ield~$f$
and its derivatives with respect~to~$x$; \item the weight of
$\vartheta$ is $[\vartheta]=-\tfrac{1}{2}$.\vspace{-2mm}
\end{itemize}
Recall that the weight of the f\/ield $f$ is not f\/ixed if
$\alpha=0$. Hence we postulate $[f]=\tfrac{3}{2}$.}

Further, we def\/ine the bosonic f\/ield $u=b+\vartheta f$ of
weight~$1=[b]=[\vartheta]+[f]$. Then we obtain the Burgers
super-f\/ield equation again! Indeed, we have
\begin{gather}\label{Burgers}
u_t=u_{xx}+uu_x.
\end{gather}
However, we note that the super-derivative
$D_\vartheta+\vartheta\,D_x$ whose square is~$D_x$ does not appear
in the reasonings. Therefore, the weights are motivated by the
background concept of supersymmetry, but they are not uniquely
def\/ined. One could easily f\/ix $[f]\mathrel{:{=}}1$ and
$\vartheta\mathrel{:{=}}0$ with the same equation~\eqref{Burgers}
in the end.

Equation~\eqref{Burgers} has the well-known recursion operator
\begin{gather}\label{RecScalarBurg}
R=D_x+\tfrac{1}{2}u+\tfrac{1}{2}u_xD_x^{-1}\quad\Longleftrightarrow
\quad\cR=U_x+\tfrac{1}{2}uU+\tfrac{1}{2}u_x\,D_x^{-1}(U).
\end{gather}
The recursion operators for the left- and right-noncommutative
$\star$-Burgers equations have been obtained by M.~G\"urses and
A.~Karasu (private communication).

Let us construct an analogue of recursion~\eqref{RecScalarBurg}
for system~\eqref{SuperBurg} if the coupling is $\alpha=0$. First
we introduce the bosonic nonlocality $v(x,t)$ of weight zero by
setting $v_x=b$. The potentials $v$ and $w$ satisfy the system
\begin{gather}\label{pBurg}
w_t=w_{xx}+v_xw_x,\qquad v_t=v_{xx}+\tfrac{1}{2}v_x^2.
\end{gather}
Next, we make a technical assumption that the bosonic variables
$v$ and $V$ of zero weight appear in the recursion
for~\eqref{SuperBurg} at most linearly. Then we f\/ind out that in
this setting there are two recursion operators of weight $1$. The
f\/irst operator,
\begin{gather*}
\cR_{[1]}^1=\binom{-\oh w B_x + \oh b_x W - \tfrac{1}{4}b_xwV +
F_x
   + \oh f_xV + \oh bF - \tfrac{1}{4}wbB - \tfrac{1}{4}fbV}%
   {2B_x+bB+b_xV},
   \end{gather*}
is the direct extension of twice the
recursion~\eqref{RecScalarBurg} for equation~\eqref{Burgers}.
Simultaneously, we obtain the shadow recursion with nonlocal
coef\/f\/icients,
\begin{gather*}
\cR_{[1]}^2=\binom{wB_x+b_xW+\oh b_xwV+2F_x+f_xV+bF+\oh wbB
   +\oh fbV +2fB}{0}.
\end{gather*}
Its dif\/ferential order is $1$; this recursion is not nilpotent.

\subsubsection[Arbitrary coupling in~(\ref{SuperBurg}): $\alpha\in\BBR$]{Arbitrary coupling
in~(\ref{SuperBurg}): $\boldsymbol{\alpha\in\BBR}$} In this
subsection we consider the case of an arbitrary non-zero real
constant~$\alpha$ in system~\eqref{SuperBurg}.

In view of the preceding subsection, we introduce the independent
variable~$\vartheta$ that anticommutes with the fermionic f\/ield
$f$ and its derivatives w.r.t.\ $x$ and such that
\begin{itemize}\vspace{-2mm}\itemsep=0pt
\item $\vartheta\cdot\vartheta=\alpha$, that is, the square
of~$\vartheta$ is now non-zero; \item the weights of the new
variable~$\vartheta$, the dimensional constant $\alpha$ (see
Remark~\ref{AxTooRestrictive}), and the fermionic f\/ield~$f$ are
related by the formulas $2[\vartheta]=[\alpha]$,
$[f]=1-[\vartheta]$.\vspace{-2mm}
\end{itemize}
This time we set $[\vartheta]=[\alpha]\mathrel{:{=}}0$,
$[f]\mathrel{:{=}}1$, and we do not consider the square root of
$D_x$ based on the variable~$\vartheta$. We emphasize
that~$\vartheta$ is not an ordinary complex number whose square
could easily be zero, negative, or positive (and $\vartheta$ would
therefore be zero, imaginary, or negative, respectively).

Again, we set $u=b+\vartheta f$; the f\/ield $u(x,t;\vartheta)$ is
homogeneous of weight~$1$. Moreover, it satisf\/ies the Burgers
equation
\begin{gather}\tag{\ref{Burgers}${}'$}\label{BurgersPrime}
u_t=u_{xx}+uu_x,
\end{gather}
but now equation~\eqref{BurgersPrime} is an equation on the
associative algebra generated by $u$ and its derivatives
w.r.t.~$x$. This is because for all $i\neq j$ we have\footnote{One
could use the notation~$\hbar$ instead of~$\alpha$.}
$[u_i,u_j]\sim\alpha\neq0$; for example,
\[
uu_x=bb_x+\vartheta^2f_xf + \vartheta\cdot(b_xf+bf_x)\neq
     bb_x-\vartheta^2f_xf + \vartheta\cdot(b_xf+bf_x)=u_xu.
\]
The geometry of equations on associative algebras has been studied
recently, see~\cite{SokolovAssociative}, by using standard notions
and computational algorithms.

Obviously, the fermionic nonlocality $w$ such that $w_x=f$ is
indif\/ferent w.r.t.~the value of the coupling constant $\alpha$.
We f\/ind out that system~\eqref{SuperBurg} always admits the
conservation law that potentiates the bosonic variable $b$. We
thus set
\[
\tilde{v}_x = b+\oh fw\alpha,\qquad \tilde{v}_t = b_x+\oh b^2+\oh
f_xw\alpha+\oh fbw\alpha.
\]
The weight of the nonlocality $\tilde{v}$ is zero. Surprisingly,
the variable $\tilde{v}$ satisf\/ies the same potential Burgers
equation as the nonlocality~$v$, see equation~\eqref{pBurg},
\begin{gather}\tag{\ref{pBurg}${}'$}\label{pBurgPrime}
w_t=w_{xx}+\tilde{v}_xw_x,\qquad
\tilde{v}_t=\tilde{v}_{xx}+\tfrac{1}{2}\tilde{v}_x^2.
\end{gather}
We conclude that equation~\eqref{pBurgPrime} potentiates
system~\eqref{SuperBurg} for all $\alpha=\vartheta\cdot\vartheta$,
although the algebraic nature of the Burgers
equation~\eqref{Burgers} w.r.t.\ the f\/ield $u=b+\vartheta f$ is
radically dif\/ferent from equation~\eqref{BurgersPrime} that
describes the Burgers f\/low on the associative algebra.

In the nonlocal setting $\{f,w\}+\{b,\tilde{v}\}$ there are two
generalizations of recursion~\eqref{RecScalarBurg}. The f\/irst
recursion of weight $1$ for equation~\eqref{BurgSystem} is
\begin{gather*}
\cR_{[1]}^1=\binom{
   b_xW+2F_x+f_x\tilde{V}+\tfrac{1}{2}f_xwW\alpha+bF+fB}%
   {\underline{2B_x+bB+b_x\tilde{V}}+\tfrac{1}{2}b_xwW\alpha
     +f_xW\alpha-fF\alpha},
\end{gather*}
here we underline the component that corresponds
to~\eqref{RecScalarBurg}. The second extension of weight~$1$ is
\begin{gather*}
\cR_{[1]}^{2,f}= -wB_x-\oh b_xw\tilde{V}+\oh
f_xwW\alpha-\tfrac{1}{4}wfbW\alpha
   -\oh wfF\alpha-\oh wbB-\oh fb\tilde{V}-fB,
\\
\cR_{[1]}^{2,b}=
\underline{2B_x+bB+b_x\tilde{V}}-wF_x\alpha+f_xW\alpha
   +\oh f_xw\tilde{V}+\oh fbW\alpha-\oh wbF\alpha\\
\phantom{\cR_{[1]}^{2,b}=}{} -\oh
wfb\tilde{V}\alpha-\tfrac{3}{2}wfB\alpha.
\end{gather*}

\begin{remark}
In the previous reasonings we treated system~\eqref{SuperBurg} as
an $N=0$ equation that involves the fermionic f\/ield but does not
contain the super-derivative~$\sd{}$. Now we enlarge the
$(x,t,f,b)$ jet space with the anticommuting independent variable
$\theta$ and the derivatives of~$f$ and~$b$ w.r.t.~$\theta$. We
have $\sd{}^2=D_x$, and the unknown functions become the
super-f\/ields $f(x,t,\theta)$ and $b(x,t,\theta)$. Physically
speaking, we permit the consideration of conservation laws at
half-integer weights for~\eqref{SuperBurg} and~\eqref{pBurgPrime}.
We discover that there are many nonlocal conservation laws for
equation~\eqref{SuperBurg}; for example, we obtain the `square
roots' of the variables $\tilde{v}$ and~$w$. We conjecture that
there are inf\/initely many $N=1$ conservation laws for
equation~\eqref{BurgSystem}. Also, there are many recursions that
involve the nonlocalities assigned to the new conservation laws
and which are nilpotent if~$\alpha=0$. This situation is analogous
to the scheme that generates non-local Hamiltonians for the $N=1$
superKdV equation~\eqref{sKdV}, see p.~\pageref{psKdVe}
and~\cite{Andrea, MathieuOpen}.
\end{remark}

\subsection[$N=2$ supersymmetric Burgers equation]{$\boldsymbol{N=2}$ supersymmetric Burgers equation}
Let us recall that $D_{\theta^i}$ and $D_x$ denote the derivatives
w.r.t.\ the independent coordinates $\theta^i$ and~$x$,
respectively, while $\sd_{i}$ are the super-derivations such that
$\sd_{i}^2=D_x$ for any~$i$.

We now admit that Axiom~8 was used when constructing the
evolutionary super-systems:
\begin{itemize}\vspace{-1mm}\itemsep=0pt
\item[8.]\label{ax8} Each of the super-derivatives
$\sd_{i}=D_{\theta^i}+\theta^iD_x$, $i=1$,\ $\ldots$,\ $N$, occurs
at least once in the r.h.s.\ of the evolutionary system if
$N\geq2$.\vspace{-1mm}
\end{itemize}

In the database~\cite{SUSY} there is a scalar, third order $N=2$
supersymmetrization of the Burgers equation; it is
\begin{gather}\label{N=2Burg}
b_t=\sd_{1}\sd_{2}b_x+bb_x, \qquad b=b(x,t,\theta^1,\theta^2),
   \qquad \sd_{i}=D_{\theta^i}+\theta^i\,D_x,\qquad i=1,2.
\end{gather}
Equation~\eqref{N=2Burg} is reduced to the second order Burgers
super-f\/ield equation $b_t=b_{xx}+bb_x$ on the super-diagonal
$\theta^1=\theta^2$, here $b=b(t,x,\theta^1,\theta^1)$.

Let us expand the super-f\/ield $b(x,t;\theta^1,\theta^2)$ in
$\theta^1$ and $\theta^2$:
\[
b=\beta(x,t)+\theta^1\xi(x,t)+\theta^2\eta(x,t)+
  \theta^1\theta^2\gamma(x,t).
\]
Hence from equation~\eqref{N=2Burg} we obtain the system for the
components of $b$:
\begin{gather}
\beta_t=-\gamma_x+\beta\beta_x, \qquad
\xi_t=\eta_{xx}+(\beta\xi)_x, \qquad
\eta_t=-\xi_{xx}+(\beta\eta)_x,\nonumber\\
\gamma_t=\beta_{xxx}+(\beta\gamma)_x-(\xi\eta)_x.
\tag{\ref{N=2Burg}${}'$}\label{N=2BurgComp}
\end{gather}
We see that the second and third equations in
system~\eqref{N=2BurgComp} are Burgers-type, that is, they contain
the dissipative terms and the remaining parts are total
derivatives. The fourth equation in~\eqref{N=2BurgComp} describing
the evolution of $\gamma$ is of KdV-type: the dispersion and two
divergent terms are present in it.

The KdV nature of the $N=2$ supersymmetric Burgers
equation~\eqref{N=2Burg} is not occasional. Indeed,
equation~\eqref{N=2Burg} is a symmetry of the $N=2$ supersymmetric
SKdV${}_4$ equation~\cite{Laberge}
\begin{gather}\label{N=2KdV4}
b_s = -b_{xxx} + \tfrac{1}{2}\bigl(b\sd_{1}\sd_{2}b\bigr)_x
   + \tfrac{3}{4}\bigl(\sd_{1}b\,\sd_{2}b\bigr)_x
   + \tfrac{3}{4}b^2b_x.
\end{gather}
Reciprocally, the SKdV${}_4$ equation~\eqref{N=2KdV4} is a higher
symmetry of the Burgers equation~\eqref{N=2Burg}, and their
bi-Hamiltonian hierarchy is of the form
\[
b_x\mapsto\text{equation~\eqref{N=2Burg}}\mapsto
\text{equation~\eqref{N=2KdV4}}\mapsto\cdots.
\]
The two equations share the recursion operator~\cite{Laberge,
MathieuOpen}. It must be noted that the relation between the
Laberge--Mathieu's $N=2$ SKdV${}_4$ equation and the $N=2$ Burgers
system was not indicated in \textit{loc.~cit}.

\section{The Boussinesq equation}\label{SecBous}
In this section we describe a super-f\/ield representation of the
dispersionless Boussinesq equation; the system at hand admits two
inf\/inite sequences of commuting symmetries of constant
dif\/ferential order ${2}$ which are generated by a weakly
non-local recursion operator of dif\/ferential order~$1$. Also, we
extend the Boussinesq system with dispersion and dissipation to
its three-parametric analogue that does not retract to it for any
values of the parameters.

\subsection{The dispersionless Boussinesq equation}
We consider the two-component system
\begin{gather}\label{DBousSyst}
f_t=b\,\sd{b},\qquad b_t=\sd{f_x}.
\end{gather}
System~\eqref{DBousSyst} is a super-representation of the
dispersionless Boussinesq equation
\begin{gather}\label{DBous}
b_{tt}=\oh(b^2)_{xx},
\end{gather}
here~$b(x,t,\theta)$ is the bosonic super-f\/ield. In what
follows, we construct a Gardner's deformation for the hydrodynamic
representation of the dispersionless Boussinesq
equation~\eqref{DBous}. Next, we transmit the properties of
Hamiltonian symmetries for equation~\eqref{DBous} onto its
supersymmetric representation~\eqref{DBousSyst}.

\subsubsection{The Gardner's deformation of the
dispersionless Boussinesq equation}\label{BosonicDBous} The
bo\-so\-nic two-component form of~\eqref{DBous} is
\begin{gather}\label{Hydro}
b_t=c_x,\qquad c_t=bb_x.
\end{gather}
Here we obviously have $c(x,t;\theta)=\sd{f(x,t;\theta)}$.
System~\eqref{Hydro} is a super-f\/ield equation of hydrodynamic
type.

\begin{remark}
Consider the dispersionless Boussinesq equation~\eqref{Hydro} with
$b(x,t)$ and $c(x,t)$. The number of independent variables in it
coincides with the number of unknown functions and equals two.
Therefore, the system at hand is linearized~\cite{Hodograph} by
using the hodograph transformation $b(x,t)$, $c(x,t)\mapsto
x(b,c)$, $t(b,c)$. Indeed, we obtain the linear autonomous system
\begin{gather}\tag{\ref{Hydro}${}'$}\label{HydroA}
x_c=t_b,\qquad t_c=\frac{x_b}{b},
\end{gather}
here $c$ is the new time and $b$ is the new spatial variable. The
solution of the hydrodynamic type system~\eqref{HydroA} is
expressed with  the Airy function.
\end{remark}

Now we construct a Gardner's deformation~\cite{PamukKale,
KuperIrish, Gardner} for system~\eqref{Hydro}. We have
\begin{gather}\tag{\ref{Hydro}${}''$}\label{HydroB}
{\binom{b}{c}}_t=\begin{pmatrix} 0 & D_x \\ D_x & 0\end{pmatrix}
{\begin{bmatrix} \delta/\delta b \\ \delta/\delta
    c\end{bmatrix}}_{(x,t)} \cH(b,c),
\end{gather}
where the density of the Hamiltonian $\cH$ is
$H=\tfrac{1}{6}b^3+\tfrac{1}{2}c^2$. We proceed with a deformation
\begin{gather}\label{HydroE}
{\binom{w^1}{w^2}}_t=\begin{pmatrix} 0 & D_x \\ D_x &
0\end{pmatrix} {\begin{bmatrix} \delta/\delta w^1 \\ \delta/\delta
w^2
    \end{bmatrix}}_{(x,t)} \bar{\cH}(w^1,w^2;\varepsilon)
\end{gather}
of equation~\eqref{HydroB}. We assume that the deformations of the
dependent variables are
\begin{subequations}
\begin{gather}
b=w^1+\varepsilon W_1(\boldsymbol{w})+\varepsilon^2
W_2(\boldsymbol{w})+
   \cdots,\qquad
c=w^2+\varepsilon \Omega_1(\boldsymbol{w})+\varepsilon^2
   \Omega_2(\boldsymbol{w})+ \cdots,\label{Homotopy}
\end{gather}
and we extend the Hamiltonian functional such that its density is
\begin{gather}
\bar{H}=H(\boldsymbol{w})+\varepsilon H_1(\boldsymbol{w}) +
   \varepsilon^2 H_2(\boldsymbol{w})+\cdots.\label{HomotopyHam}
\end{gather}
\end{subequations}
We also expand the f\/ields $w^1(x,t;\varepsilon)$ and
$w^2(x,t;\varepsilon)$ in $\varepsilon$:
\[
w^i(x,t;\varepsilon)=\sum_{k=0}^{+\infty} w^i_k\,\varepsilon^k,
\qquad i=1,2.
\]
Recall that equation~\eqref{Hydro} is in the divergent form.
Therefore, the Taylor coef\/f\/icients $w^i_k$ are termwise
conserved, and from~\eqref{Homotopy} we get the initial conditions
$w^1_0=b$, $w^2_0=c$ and the recurrence relations for the
conserved densities.

Now we truncate expansions~\eqref{Homotopy}, \eqref{HomotopyHam}
to polynomials of suf\/f\/iciently large degrees. By using the
homogeneity reasonings, in view of $[\varepsilon]=-3$, we then
reduce the deformation problem to an algebraic system for the
undetermined coef\/f\/icients in these expansions. We solve the
algebraic system by using the program \textsc{Crack}
\cite{WolfCrack} and f\/inally obtain the deformation
\begin{gather*}
b=w^1+\varepsilon w^1w^2,\\
c=w^2+\tfrac{1}{3}\varepsilon\bigl(w^1\bigr)^3
   +\varepsilon\bigl(w^2\bigr)^2
   +\tfrac{1}{3}\varepsilon^2\bigl(w^2\bigr)^3,\\
\bar{H}=\tfrac{1}{6}\bigl(w^1\bigr)^3
   +\tfrac{1}{2}\bigl(w^2\bigr)^2
   +\tfrac{1}{6}\varepsilon\bigl(w^2\bigr)^3.
\end{gather*}
The initial terms in the two sequences of the conserved densities
are
\begin{alignat}{5}
& w^1_0=b,\qquad && w^1_1=-bc,\qquad & & w^1_2=2bc^2+\tfrac{1}{3}b^4,\quad & &\ldots,&\nonumber\\
& w^2_0=c,\qquad && w^2_1=-c^2-\tfrac{1}{3}b^3,\qquad &&
w^2_2=\tfrac{5}{3}c^3+\tfrac{5}{3}b^3c,\quad & &\ldots.
&\label{DBousHams}
\end{alignat}
Taking into account that the Hamiltonian operator
$\left(\begin{smallmatrix} 0 & D_x \\ D_x &
0\end{smallmatrix}\right)$ maps gradients of conservation laws for
equation~\eqref{Hydro} to its symmetries, see~\cite{JKKersten}, we
thus conclude that~\eqref{DBousHams} describes two inf\/inite
sequences of Hamiltonians for~\eqref{Hydro}. They determine
inf\/initely many contact symmetry f\/lows, which are local
w.r.t.\ $b$ and~$c$. Moreover, each f\/low lies in the image of
$D_x$ by construction, and hence they induce local symmetry
transformations of the super-variable $f(x,t;\theta)$ such that
$c=\sd{f}$. We emphasize that we did not even need a recursion
operator for~\eqref{Hydro} to obtain the contact symmetry f\/lows
and prove their locality.

\begin{remark}
The problem of constructing integrable deformations for
homogeneous (super-) PDE is another practical application of
the~\textsc{Crack} solver~\cite{WolfCrack} for large
overdetermined systems of~nonlinear algebraic equation. This
application has not been previously considered within the
framework of~\cite{Dubna, WolfCrack, SUSY}.
\end{remark}

\subsubsection[Supersymmetric representation of equation~(\ref{DBous})]{Supersymmetric
representation of equation~(\ref{DBous})}

Now we discuss the Hamiltonian properties of
system~\eqref{DBousSyst}. It is homogeneous w.r.t.\ multiply
def\/ined weights; we let the primary set be $[f]=[b]=1$,
$[t]=-1\oh$, $[x]=-1$. With respect to this set,
system~\eqref{DBousSyst} admits two inf\/inite sequences of
Hamiltonian symmetry f\/lows of unbounded weights $-1$, $-1\oh$,
$-2\oh$, $-3$, $\ldots$, and constant dif\/ferential order $2$.
These two sequences start with the f\/lows
\begin{gather}\label{SymDBousSyst}
\binom{f_x}{b_x} \mapsto
\binom{\sd{b}\,b^2+\sd{f}\,f_x}{\sd{f_x}\,b+\sd{f}\,b_x}\mapsto\cdots,
\qquad \binom{f_t}{b_t}\mapsto \binom{\sd{f}\,\sd{b}\,b+\oh
f_xb^2}{\sd{f_x}\,\sd{f}\,+\oh b_xb^2} \mapsto\cdots.
\end{gather}
The skew-adjoint operator $A=\smash{\left(\begin{smallmatrix}0&\sd
\\ -\sd&0\end{smallmatrix}\right)}$ is a Hamiltonian structure for
the symmetries in~\eqref{SymDBousSyst}, and, by
Section~\ref{BosonicDBous}, all the f\/lows possess the
Hamiltonian functionals. There are two Casimirs $H_0^{(1)}=b$,
$H_0^{(2)}=\sd{f}$ for equation~\eqref{DBousSyst}.
From~\eqref{DBousHams} we obtain the Hamiltonians with densities
\begin{alignat}{4}
& H_1^{(1)}=b\sd{f}, &&
  H_2^{(1)}=\tfrac{1}{12}b^4+\tfrac{1}{2}b(\sd{f})^2, && \ldots,&\nonumber\\
& H_1^{(2)}=\tfrac{1}{2}(\sd{f})^2+\tfrac{1}{6}b^3,\qquad &&
  H_2^{(2)}=\tfrac{1}{6}(\sd{f})^3+\tfrac{1}{6}b^3\sd{f},\quad && \ldots.&\tag{\ref{DBousHams}${}'$}\label{DBousHam}
\end{alignat}

Let us construct the nonlocal recursion that maps the symmetries
in~\eqref{SymDBousSyst}. To this end, we introduce the nonlocality
$w$ such that $w_t=b^2/2$ and $\sd{w}=f$. We further let the
variable $v$ be such that $v_t=\sd{f}$, $v_x=b$.
Then we obtain the nonlocal recursion 
\begin{gather}\label{RecDBousSyst}
\cR_{[1\frac{1}{2}]}=
  \binom{\sd{b}\,b\,V + \oh b^2\,\sd{V}+\tfrac{3}{4}\sd{f}\,F+
    \tfrac{3}{4}f_x\,W}{
    \sd{f_x}\,V + \oh b\,\underline{\sd{F}}
     +\tfrac{3}{4}\sd{f}\,B+\tfrac{3}{4}b_x\,W}
\end{gather}
of dif\/ferential order~$1$. Indeed, the order of the f\/low
$b_{s_{i+1}}$ equals the order of $f_{s_{i}}$ plus~$1$ owing to
the presence of the underlined dif\/ferential
operator~$\sd{}=D_\theta+\theta D_x$ in~$\cR^b$.

\medskip

\noindent {\bf Proposition.} {\it Recursion~\eqref{RecDBousSyst}
proliferates the symmetries of constant differential order~$2$ for
the dispersionless Boussinesq super-field
equation~\eqref{DBousSyst}. The initial terms of the symmetry
sequences are given in~\eqref{SymDBousSyst}, and their
Hamiltonians are~\eqref{DBousHam}.}

\medskip

The assertion follows from the results of
Section~\ref{BosonicDBous}.

We note that Hamiltonian symmetries~\eqref{SymDBousSyst} are an
example of inf\/initely many f\/lows that are not obtained by a
recurrence multiplication scheme, see p.~\pageref{DefRecRelation}
for def\/inition, although their dif\/ferential order is constant.
Indeed, they are contact in the coordinates $b$, $c$ and are of
second order w.r.t.~$f$ and~$b$.

\subsection{The Boussinesq equation with dispersion and dissipation}
Representation~\eqref{DBousSyst} of the dispersionless Boussinesq
super-f\/ield equation is embedded in the one-parametric family of
supersymmetric systems
\begin{subequations}\label{BousParamBoth}
\begin{gather}\label{BousParam}
f_t=b\sd{b}+\sd{b_{xx}}-\alpha f_{xx},\qquad b_t=\sd{f_x}+\alpha
b_{xx}, \qquad \alpha\in\BBR.
\end{gather}
By def\/inition, put $c=\sd{f}$. Then from~\eqref{BousParam} we
get the bosonic Boussinesq system (see, e.g.,~\cite{PamukKale,
KuperIrish, Olver})
\begin{gather}
c_t=bb_x+b_{xxx}-\alpha c_{xx}, \qquad b_t=c_x+\alpha b_{xx},
\qquad \alpha \in\BBR.\label{BousParamEven}
\end{gather}
\end{subequations}
The two systems~\eqref{BousParamBoth} are homogeneous w.r.t.\ the
weights $[b]=2$, $[f]=2\oh$, $[c]=3$, and $[t]=-2$. We see that
equation~\eqref{BousParam} has no nontrivial bosonic limit.
Indeed, one can not set $f\equiv0$ such that the fermionic
equation remains consistent unless~$b=\const$.

If $\alpha=0$, then~\eqref{BousParamBoth} is the Boussinesq
equation with dispersion. The terms involving $\alpha$ describe
the dissipation. There is a well-known recursion operator of
weight $[s_R]=-3$ for the Boussinesq equation without
dissipation~\cite{Olver}. The recursion for the fermionic
component in~\eqref{BousParam} is then, roughly speaking, the
component corresponding to $c$ in the recursion
for~\eqref{BousParamEven} conjugated by~$\sd{}$.

If $\alpha$ is non-zero, then system~\eqref{BousParamEven} is not
reduced to a scalar fourth order super-f\/ield equation. In this
case, the system is translation invariant and for all
$\alpha\in\BBR$ it admits symmetries of weights
$[s]=-\bigl(3k+\tfrac{3}{2}\pm\tfrac{1}{2}\bigr)$, $k\in\BBN$,
$k\geq0$.

\subsection{The multi-parametric Boussinesq-type equation}
Finally, we construct the Boussinesq-type system using
representation~\eqref{BousParam} for the Boussinesq equation. From
the database~\cite{SUSY} we obtain the system
\begin{gather}
f_t=\alpha\beta fb
-\alpha\gamma b\sd{b}-\gamma^2\sd{b_{xx}}-\beta\gamma f_{xx},\nonumber\\
b_t=\alpha\beta b^2+\beta^2\sd{f_x}+\beta\gamma
b_{xx},\label{BousEmbed}
\end{gather}
here $\alpha$, $\beta$, $\gamma\in\BBR$. This system is an
analogue of Boussinesq equation~\eqref{BousParamBoth} but does not
retract to it for any value of the parameters. Similarly to
equation~\eqref{BousParam}, system~\eqref{BousEmbed} does not have
a~nontrivial bosonic limit at $f\equiv0$.

The translation invariance of equation~\eqref{BousEmbed} is
obvious; we recall that $[x]=-1$ and $[t]=-2$. Also,
equation~\eqref{BousEmbed} admits the symmetry
\begin{gather*}
{f}_s = -\sd{b}b_{xx}\gamma^3 + \sd{b_x}b_x\gamma^3 +
\sd{b_x}\sd{f}\beta\gamma^2  - \sd{f_x}\sd{b}\beta\gamma^2
-\sd{f_x}f\beta^2\gamma + \sd{f}f_x\beta^2\gamma \\
\phantom{{f}_s =}{}
 - b_{xx}f\beta\gamma^2
+ b_xf_x\beta\gamma^2,
\\
{b}_s = -\sd{b}f_x\beta^2\gamma + \sd{b_x}f\beta^2\gamma +
\sd{b_x}\sd{b}\beta\gamma^2 + f_xf\beta^3
\end{gather*}
such that $[s]=-4$, and there is the symmetry
\begin{gather*}
{f}_{\bar{s}} =  \sd{b}f_x\beta\gamma^2 - \sd{b_x}f\beta\gamma^2 -
\sd{b_x}\sd{b}\gamma^3 - b_x^2\gamma^3 - (\sd{f})^2\beta^2\gamma
 - f_xf\beta^2\gamma - 2\sd{f}b_x\beta\gamma^2,
\\
b_{\bar{s}} = \sd{b}b_x\beta\gamma^2 + \sd{f}\sd{b}\beta^2\gamma +
\sd{f}f\beta^2\gamma + b_xf\beta^2\gamma,
\end{gather*}
here the weight of the odd parameter $\bar{s}$ is
$[\bar{s}]=-3\oh$. No more symmetries exist for
equation~\eqref{BousEmbed} if $[s]\geq-7$ and if the parity of~$s$
is arbitrary. Quite strangely, there are no conservation laws for
equation~\eqref{BousEmbed} within the weights not greater
than~$11$; hence no nonlocalities were constructed and no
recursion is currently known for system~\eqref{BousEmbed}.

\section{Conclusion}
In this paper, we investigated the integrability of fermionic
extensions and supersymmetric generalizations for the KdV,
Burgers, and Boussinesq equations. Recursion operators for their
symmetry algebras were obtained. Also, we analyzed the properties
of Gardner's deformations for the dispersionless Boussinesq
equation~\eqref{Hydro} and $N=1$ supersymmetric KdV
equation~\eqref{sKdV}.

The boson$+$fermion super-f\/ield
representations~\eqref{BurgSystem}, \eqref{DBousSyst},
and~\eqref{BousParamBoth} for the Burgers and Boussinesq equations
are remarkable by themselves. Indeed, the roles of the independent
variables $x$ and $t$ are swapped in system~\eqref{BurgSystem}:
$x$ is the time and $t$ is the spatial coordinate in
equation~\eqref{BurgersInverse}. The dispersionless Boussinesq
system~\eqref{DBousSyst} admits two recursive
sequences~\eqref{SymDBousSyst} of Hamiltonian symmetries whose
dif\/ferential orders is constant, while the dif\/ferential order
of nonlocal recursion~\eqref{RecDBousSyst} is strictly positive.
The Boussinesq-type system~\eqref{BousEmbed} is multi-parametric
and contains the Boussinesq equation with dispersion as a
component but can not be reduced to it at any values of the
parameters. We f\/inally note that all the Boussinesq-type systems
\eqref{DBousSyst}, \eqref{BousParam}, \eqref{BousEmbed}, as well
as representation~\eqref{BurgSystem} for the Burgers equation, do
not have bosonic limits at~$f\equiv0$.

Using fermionic extension~\eqref{SuperBurg} of the Burgers
equation, we conclude that the `direct $N\geq1$
supersymmetrization' \cite{Mathieu} based on replacing the
derivatives with super-derivatives in a PDE is not a unique way to
obtain its generalizations. Indeed, the new systems can contain
the fermionic f\/ields but no super-derivatives. Hence we indicate
three possible types of these generalizations.
\begin{enumerate}\vspace{-2mm}\itemsep=0pt
\item The new super-f\/ields combine the old f\/ields with the new
added components and satisfy manifestly $N\geq1$ supersymmetric
equations, see the superKdV equation~\eqref{sKdV}, for example.
\item The evolution of new fermionic f\/ields $f(x,t)$ is coupled
with the original equation, see~\eqref{SuperBurg}. Then, new
independent (Grassmann, or super\/-)\/variables $\vartheta$ are
introduced such that the extended system is reduced to a smaller
equation. The geometric structures of the extension are then
inherited from the f\/inal system by routine expansions in the new
variables~$\vartheta$, see~\eqref{Burgers}. \item Similarly to the
previous case, the new f\/ield is constructed by using a new
independent variable that anticommutes with the fermionic function
but not with itself. Then the resulting system is an evolution
equation on an associative algebra~\cite{SokolovAssociative}, see
equation~\eqref{BurgersPrime} on
p.~\pageref{BurgersPrime}.\vspace{-2mm}
\end{enumerate}
We conclude that the equations on associative algebras present a
nontrivial way to generalize coupled fermion$+$boson systems.

\subsection*{Acknowledgements}
The authors thank V.V.~Sokolov for formulation of the
classif\/ication problem and also thank A.V.~Mikhailov,
A.~Sergyeyev, and A.S.~Sorin for helpful discussions. The authors
are greatly indebted to the referees for their remarks and
suggestions. T.W.\ wishes to thank W.~Neun for discussions and the
SHARCNET\ consortia for computer access. The research of A.K.\ was
partially supported by the Scientif\/ic and Technological Research
Council of Turkey (TUBITAK). A part of this research was carried
out while A.K.\ was visiting at Brock University and at Middle
East Technical University (Ankara).

\LastPageEnding
\end{document}